%% file: Genptw.tex
\documentclass[letterpaper]{article} % DO NOT CHANGE THIS
\usepackage{aaai2026}  % DO NOT CHANGE THIS
\usepackage{times}  % DO NOT CHANGE THIS
\usepackage{helvet}  % DO NOT CHANGE THIS
\usepackage{courier}  % DO NOT CHANGE THIS
\usepackage[hyphens]{url}  % DO NOT CHANGE THIS
\usepackage{graphicx} % DO NOT CHANGE THIS
\usepackage{natbib}  % DO NOT CHANGE THIS AND DO NOT ADD ANY OPTIONS TO IT
\usepackage{caption} % DO NOT CHANGE THIS AND DO NOT ADD ANY OPTIONS TO IT
\usepackage{amsfonts} 
\usepackage{algorithm}
\usepackage{algorithmic}
\usepackage{amsmath} % Added to support \text in math mode
\usepackage{newfloat}
\usepackage{listings}
\usepackage{multicol}
\usepackage{multirow}
\usepackage{tabularx}
\usepackage{makecell}
\usepackage{color}
\usepackage[table]{xcolor}
\usepackage{arydshln} 
\usepackage{colortbl}
\usepackage{booktabs}
\usepackage{pifont}

\DeclareCaptionStyle{ruled}{labelfont=normalfont,labelsep=colon,strut=off} % DO NOT CHANGE THIS
\floatstyle{ruled}
\newfloat{listing}{tb}{lst}{}
\floatname{listing}{Listing}

\usepackage{multirow}

\title{GenPTW: Latent Image Watermarking for Provenance Tracing and Tamper Localization}
\author {
    Zhenliang Gan\textsuperscript{\rm 1},
    Chunya Liu\textsuperscript{\rm 2},\\
    Yichao Tang\textsuperscript{\rm 1},
    Binghao Wang\textsuperscript{\rm 2},
    Shiwen Cui\textsuperscript{\rm 2},
    Weiqiang Wang\textsuperscript{\rm 2},
    Xinpeng Zhang\textsuperscript{\rm 1}\thanks{Corresponding Author.}
}
\affiliations {
    \textsuperscript{\rm 1}{College of Computer Science and Artificial Intelligence, Fudan University, China}\\
    \textsuperscript{\rm 2}Ant Group, China\\
    zlgan23@m.fudan.edu.cn,
    \{liuchunya.lcy, weiqiang.wwq, binghao.wbh, donn.csw\}@antgroup.com,\\
    \{yichao\_tang, zhangxinpeng\}@fudan.edu.cn
}
\usepackage{bibentry}
\begin{document}

\maketitle

\begin{abstract}
% 生成图像模型的迅猛发展带来了 AI-generated content (AIGC) creation的新机遇，也引发了内容归属模糊与篡改检测困难等安全挑战。
% 现有水印方法普遍缺乏篡改定位能力，或仅适用于生成后的后处理场景，难以满足实际取证需求。
% 为解决上述问题，我们提出了GenPTW：一个通用的潜空间图像水印框架，兼容生成时嵌入与生成后嵌入两种范式，统一实现生成追溯（Provenance Tracing）与篡改定位（Tamper Localization）任务。GenPTW不影响原始生成流程，适配LDMs和VAR，是plug-and-play.
% 嵌入：我们设计了跨注意力融合模块（CAF），根据潜空间特征自适应选择最优水印嵌入策略；并引入空间融合模块（SF），将完整水印信息注入潜特征以实现高精度篡改定位。
% 提取：我们设计了Tamper-Aware Extractor，将两个任务紧密耦合，以提升在复杂编辑场景下的鲁棒性与定位精度；
% 大量实验证明，GenPTW achieves high image fidelity, flexible forensics, and strong robustness against complex AIGC edits.

The proliferation of generative image models has revolutionized AIGC creation while amplifying concerns over content provenance and manipulation forensics.
Existing methods are typically either unable to localize tampering or restricted to specific generative settings, limiting their practical utility.
We propose \textbf{GenPTW}, a \textbf{Gen}eral watermarking framework that unifies \textbf{P}rovenance tracing and \textbf{T}amper localization in latent space.
It supports both in-generation and post-generation embedding without altering the generative process, and is plug-and-play compatible with latent diffusion models (LDMs) and visual autoregressive (VAR) models.
% To enable precise tracing and tamper localization, we propose a dual-module design: a cross-attention fusion mechanism adaptively embeds watermark guided by latent features, while a spatial fusion module reinforces localization by injecting complete watermark information.
To achieve precise provenance tracing and tamper localization, we embed the watermark using two complementary mechanisms: cross-attention fusion aligned with latent semantics and spatial fusion providing explicit spatial guidance for edit sensitivity.
A tamper-aware extractor jointly conducts provenance tracing and tamper localization by leveraging watermark features together with high-frequency features.
% A tamper-aware extractor further unifies provenance and manipulation decoding, tightly coupling watermark semantics with forensic objectives.
Experiments show that GenPTW maintains high visual fidelity and strong robustness against diverse AIGC-editing.
\end{abstract}

\input{sec/1-intro}
\input{sec/2-releat}
\input{sec/3-method}

\input{sec/4-exper}

\section{Conclusion}
%在本文中，我们提出了GenPTW，一种微调隐空间变量的内生水印方法。据我们所了解，是第一个支持版权溯源和篡改检测定位的内生水印方案。我们通过水印信息模块和水印特征模块，将水印与隐空间变量z 在空域和高维特征空间进行充分融合。 在训练阶段，to overcome gen quality，leverage JND显性最小化水印对图片的影响；to improve robust 通过DCT频谱转换分离低频和高频信息，低频信息用于鲁棒水印的提取，while高频信息用于脆弱水印的篡改定位。，同时采用mini-batch策略处理真实AIGC inpainting，全图VAE重构和替换篡改的攻击； 。实验表明，我们的方案中低频鲁棒水印能够应对任意setting下aigc的攻击，高频脆弱水印可以应对任意局部编辑操作后的篡改定位。
In this paper, we propose \textbf{\textit{GenPTW}}, a \textbf{Gen}eral latent-space watermarking framework for proactive \textbf{P}rovenance tracing and \textbf{T}amper localization.
To the best of our knowledge, GenPTW is the first framework to unify these two capabilities through a single embedding in latent space, seamlessly supporting both in-generation and post-hoc usage.
To achieve this, We propose a novel strategy that adapts the watermark to image latent space, incorporating both a cross-attention fusion module and a spatial fusion module. 
The cross-attention fusion module embeds the watermark based on latent features, while the spatial fusion module integrates the watermark as spatial cues into final features. 
Furthermore, we introduce a tamper-aware extractor that combines watermark signals with high-frequency features to enable precise tamper localization.
% Extensive experiments demonstrate that GenPTW consistently outperforms existing watermarking and tamper location baselines in terms of fidelity, localization precision, and robustness under diverse tampering scenarios.
Extensive experiments demonstrate that GenPTW consistently outperforms SOTA watermarking and forensic baselines in fidelity, localization accuracy, and robustness across diverse manipulation scenarios.

% In this paper, we propose GenPTW, a in-generation watermarking method that fine-tune latent space z variables. 
% To out knowledge, it is the first in-generation watermarking solution that supports provenance tracing and tamper localization. 
% We integrate the watermark with the latent z variable of image in both spatial domain and high-dimensional feature space by using watermark information process module and watermark feature module.
% During the training phase, we improve generation quality by leveraging JND as loss weight to minimize the watermark's visible impact on images. 
% To enhance robustness, we adopt a mini-batch strategy to handle real AIGC inpainting, full-image VAE reconstruction, and other copy-move tampering attacks. Additionally, we use DCT spectrum conversion to split low-frequency and high-frequency signal. Low-frequency info is used for extracting robust watermarks, while high-frequency info is used for fragile watermarks tamper localization.

% Experiments demonstrate that our approach's low-frequency robust watermarks can withstand attacks from any AIGC setting. Meanwhile, the high-frequency fragile watermarks effectively locate tampering after any local editing operations.

\section{Acknowledgments}
% This work was supported by the National Natural Science Foundation of China (Grants No. U22B2047, 62450067 and 62502093). 
This work was supported by the National Natural Science Foundation of China (Grants No. 62450067, U22B2047 and 62502093). 
The authors from Ant Group are supported by the Leading Innovative and Entrepreneur Team Introduction Program of Hangzhou (Grant No. TD2022005).

\bibliography{aaai2026}

\end{document}

%% file: sec/1-intro.tex
\section{Introduction}
\label{sec_intro}
% 生成模型正以前所未有的速度演进，尤其是文本到图像（Text-to-Image, T2I）扩散模型如 Stable Diffusion、DALL·E 3 和 Imagen；还有近期展现出巨大潜力的视觉自回归模型（Vision AutoRegressive, VAR）。
% 他们具备生成高度逼真、视觉冲击力强的图像的能力，并可以灵活编辑图像，正在重塑视觉内容的生产与传播格局。
% 然而，这种令人惊艳的生成和编辑能力也如双刃剑，带来了内容滥用、版权归属模糊与篡改检测困难等一系列安全风险。
Generative models are advancing at an unprecedented pace, particularly text-to-image diffusion models such as Stable Diffusion, DALL·E 3, and Imagen, as well as the emerging VAR models that demonstrate remarkable potential. These models are capable of synthesizing highly realistic and visually striking images with flexible editability, reshaping the paradigm of visual content creation and dissemination. However, such impressive generative and editing capabilities also pose a double-edged sword, introducing security risks such as content misuse, ambiguous ownership, and challenges in tamper detection.These risks underscore two core challenges: attributing content ownership and detecting potential manipulations.
% 近年来，图像被盗用、篡改，甚至伪造成虚假证据的事件屡见不鲜，正在侵蚀公共舆论环境与法律信任机制。
% 这类问题本质上揭示了两个核心挑战：内容溯源与内容完整性验证。
% 即，一：是谁生成/创作的，属于谁
% 二，有没有被篡改，哪里被改动了
% Image misuse, tampering, and falsification are increasingly common, eroding public trust and legal integrity.
% These risks underscore two core challenges: attributing content ownership and detecting potential manipulations.

\begin{figure}[t]
    \centering
    \includegraphics[width=1.0\linewidth]{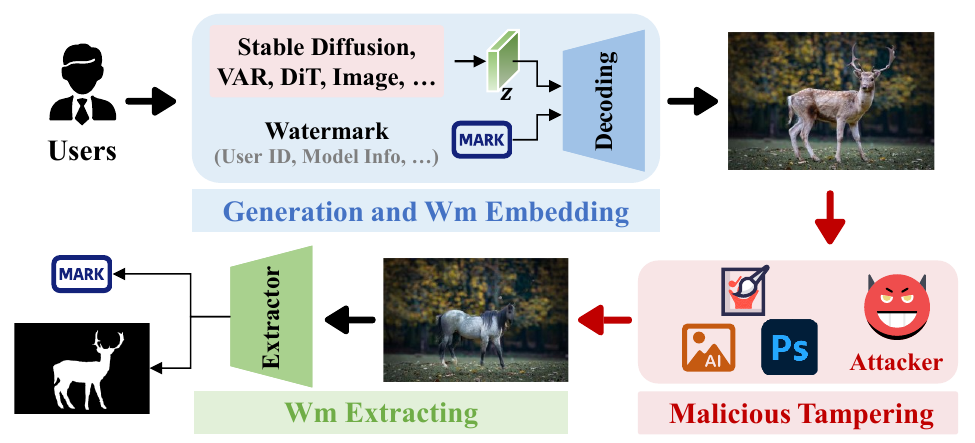}
    \caption{The process of embedding and extracting \textbf{GenPTW} for dual forensic objectives.}
    \label{fig:intro}
    % \vspace{-3mm}
\end{figure}
% 图像水印技术被广泛应用于source tracing and copyright protection。
% 然而现有方法大多聚焦于ownership identification，难以支持对篡改区域的精准定位。
% 而篡改定位对于厘清原始生成与后续编辑的边界、明确模型责任、判断篡改意图等，具有不可或缺的价值，是实现图像全面可追溯的关键环节。
% Image watermarking has been widely adopted for source tracing and copyright protection. Yet most existing methods concentrate solely on ownership identification, offering limited support for precise tamper localization. 
% However, localization is indispensable for delineating the boundary between generation and editing, attributing responsibility to the originating model, and inferring editing intent—essential components of comprehensive forensic traceability.
Image watermarking has been widely adopted for source tracing and copyright protection.
However, most existing methods remain centered on ownership identification and offer limited support for tamper localization. 
Accurate localization is essential for distinguishing generated content from subsequent edits, assigning responsibility to the originating model, and understanding the intent behind manipulation. These functions together underpin a comprehensive framework for forensic traceability.
% 为此，一些研究开始探索将水印溯源与篡改检测相结合。如 SepMark~\cite{wu2023sepmark} introduced the first deep separable watermark for robust copyright protection and deepfake detection.，EditGuard~\cite{zhang2024editguard} 利用图像隐写中的局部脆弱性实现篡改区域定位。
% 但这些方法均属于“生成后处理”范式，即图像生成完成后再进行水印嵌入，但这些方法存在与生成流程割裂、部署复杂、效率低下等问题。
Recent studies have begun to unify provenance tracing and tamper localization. For example, \emph{SepMark}~\cite{wu2023sepmark} introduces a separable watermark architecture for robust ownership verification and deepfake detection, while \emph{EditGuard}~\cite{zhang2024editguard} exploits spatial fragility in steganography to identify tampered regions.
A common limitation of these approaches is their reliance on post-generation embedding, where watermarks are added only after image synthesis. This design decouples watermarking from the generative process, complicates deployment, and reduces overall efficiency.

% 为此，近期研究逐步转向“生成时嵌入”（In-Generation）的水印范式，即在图像生成过程中直接注入水印。
% 但多数方法①缺乏篡改区域定位能力，②无法抵抗 AIGC 编辑。限制了其取证效果与实用性。
% 此外，大多针对扩散模型设计，没有一种通用的水印架构设计模型同时支撑扩散模型和VAR还有已有图像的后处理情况。
To address these limitations, recent research has shifted toward an \textit{in-generation} paradigm, injecting watermarks directly during image generation. Despite this advancement, most existing approaches still lack tamper localization capabilities and remain fragile under downstream AIGC editing, thus limiting their forensic reliability and real-world applicability.
Moreover, current designs are largely confined to diffusion models, lacking a unified watermarking framework that can generalize across LDMs, VAR models, and post hoc scenarios involving pre-generated images.
% 目前，大部分图像生成模型（潜扩散模型LDMs，视觉自回归VAR）都在latent space进行生成，然后解码到图像。
% 因此，为解决上述问题，我们提出了 **GenPTW**，一种在潜空间嵌入水印的general图像水印框架，不影响原始生成流程，同时支持潜扩散模型和VAR，并可以拓展至Post-hoc情况。解决如图 1 所示的“双重取证任务”：
% - **生成追溯（Provenance Tracing）**：图像追溯其来源与归属；
% - **篡改定位（Tamper Localization）**：精确标注图像中被篡改的区域。
Most image generation models, including LDMs and VAR models, operate in the latent space and decode to images afterward.
To address limitations in forensic capacity and model generalization, we propose GenPTW, a general-purpose image watermarking framework that embeds watermarks in the latent space. GenPTW preserves the original generation pipeline and is compatible with LDMs, VAR, and post hoc processing of existing images.
As shown in Fig.~\ref{fig:intro}, GenPTW targets a dual forensic objective:
\textbf{(1) Provenance tracing}: Identifying the source and ownership of images.
\textbf{(2) Tamper Localization:} Marking the regions that have been manipulated.

% 为实现对原始生成流程和原始图像的影响最小，并可以篡改定位，我们引入以下两个模块。
% 首先，我们设计了水印和潜空间交互的\textbf{跨注意力融合Module（Cross-Attention Fusion，CAF）}，由当前潜空间特征动态选择最佳水印嵌入策略，在保证鲁棒性的同时最小化视觉失真；
% 其次，（Spatial Fusion，SF）模块，将初始的水印特征逐点广播到潜空间特征upsample的最后一层。每一个点有一个完整的水印信息，将空间信息注入图像的潜特征，实现高精准度的篡改定位。
% Furthermore, a gradient-guided encoder is employed to embed the watermark under Just Noticeable Difference (JND) constraints, using a modification cost map, and is regularized across multiple latent-space scales to ensure both invisibility and fidelity.
% 为了将Provenance Tracing和Tamper Localization  integrate into a unified framework，我们设计了一个xxx提取器。水印提取和篡改定位共享特征提取器，实现编码与解码过程的高度耦合；
To minimally affect the generative process while enabling precise tamper localization, we introduce two dedicated modules. 
First, the Cross-Attention Fusion (CAF) module dynamically selects optimal watermark embedding strategies conditioned on latent features, balancing robustness and imperceptibility. 
Second, the Spatial Fusion (SF) module expands the watermark into a spatial guidance map and injects it into the final upsampled latent features of the decoder to enhance tamper localization.
A gradient-guided encoder further embeds the watermark under Just Noticeable Difference (JND) constraints, guided by a modification cost map and regularized across multi-scale latent features. 
Finally, a Tamper-Aware Extractor tightly integrates provenance tracing and tamper localization through a unified feature backbone, ensuring effective and robust decoding.
{Our main contributions are summarized as follows:}
\begin{itemize}
% \item We design a novel embedding strategy employing Cross-Attention and Spatial Fusion modules to achieve adaptive watermarking. The cross-attention fusion module embeds the watermark based on latent features, while the spatial fusion module incorporates the watermark as spatial hints into final features. This strategy improves imperceptibility of image, robustness of watermark extraction, and precision of tampering localization.
% \item We develop Cross-Attention Fusion(CAF) module that adaptively embeds watermark based on latent semantics, balancing robustness and imperceptibility.
% \item We design Spatial Fusion(SF) module to inject watermark as spatial hints into the final upsampled features of the decoder, and design a Tamper-Aware Extractor to integrate watermark cues and high-frequency features for precise manipulation localization.

% \item Extensive experiments show that GenPTW supports both in-generation and post-generation watermarking, delivering high fidelity, robustness, and localization accuracy under diverse AIGC-editing.
% \item 为了鲁棒提取水印Furthermore, we introduce a tamper-aware extractor that combines watermark signals with high-frequency features to enable precise tamper localization.
% \item We propose GenPTW, a general latent-space watermarking framework supporting both in-generation and post-hoc paradigms, unifying provenance tracing and tamper localization.
\item We propose \textbf{GenPTW}, the first framework to unify provenance tracing and tamper localization through a single embedding in latent space, seamlessly supporting both in-generation and post-hoc usage.

\item We design a novel embedding strategy employing Cross-Attention and Spatial Fusion modules to achieve adaptive watermarking, that improves imperceptibility of watermark, robustness of watermark extraction, and precision of tampering localization.

\item We introduce a tamper-aware extractor that jointly leverages embedded watermark cues and high-frequency features to enable robust watermark decoding and accurate tamper localization, even under severe degradations.

\item Extensive experiments show that GenPTW achieves superior performance over existing watermarking and forensic baselines in terms of visual fidelity, flexibility, and robustness.
\end{itemize}

%% file: sec/2-releat.tex
\section{Related Work}
\subsection{Image Tamper Localization}
% Detection and localization of image tampering is a critical task in digital media forensics, mainly categorized into passive and proactive methods.  Passive methods are applicable to a wider range of scenarios, while proactive methods is more effective in specific application scenarios. 
% \subsubsection{}
Localization of image tampering is a critical task in digital media forensics, mainly categorized into passive and proactive methods. \textbf{\emph{Passive  methods}} examine intrinsic attributes such as statistical features, lighting conditions, color distribution, noise discrepancies, and DCT correlations \cite{chen2021image, wu2018image, islam2020doa, zhuang2021image, guillaro2023trufor, yu2024diffforensics} to identify tampering without external information. But they require real domain data for optimal performance and lack generalization ability. \textbf{\emph{Proactive  methods}} involve embedding imperceptible markers or watermarks into images, which are easily destroyed or altered when tampering occurs. Traditional fragile watermarking method, such as block-wise hash verification or pixel-level grayscale analysis \cite{cheng2012refining, lin2023fragile, nr2022fragile, hurrah2019dual}, have limited localization accuracy and flexibility. To address these limitations, deep learning-based approaches\cite{wang2021faketagger, asnani2023malp} have been developed. 
More recently, methods like EditGuard \cite{zhang2024editguard} and OmniGuard \cite{Zhang_2025_CVPR} have employed two-stage embedding for pixel-level localization and copyright protection, through combining steganography and watermark technology. However, they still require Pre-defined template to ensure precise pixel-level tamper localization.
% 图像篡改定位 被动 主动
% 主动
%  Imuge [58, 60] adopted the self-embedding
%  mechanism and an efficient attack layer to realize tamper
%  localization and self-recovery. 
%  Draw [21] embedded wa
% termarks in RAW images to enhance passive tamper local
% ization networks’ resilience to lossy operations like JPEG
%  compression, blurring, and re-scaling. 

% Moreover, existing endogenous watermarking methods are vulnerable to AIGC edits and aggressive degradations (e.g., composite attacks, JPEG compression), often resulting in complete watermark loss. Most of these methods also lack the capability to localize tampered regions, restricting their forensic effectiveness.
\subsection{Post-hoc Image Watermarking}

% \subsubsection{Post-hoc Watermarking}
% \textbf{\emph{Post-hoc Watermarking.}}
Post-hoc watermarking methods embed watermarks into already existing images, and encompass both traditional and deep learning-based approaches. 
% Traditional methods such as DwtDct \cite{navas2008dwt} and DwtDctSvd \cite{navas2008dwt}, which design embedding mechanisms in imperceptible spatial or frequency domains for watermark embedding.
% Traditional methods such as Dct\cite{chu2003dct} and DwtDctSvd\cite{navas2008dwt}, which design embedding mechanisms in imperceptible spatial or frequency domains for watermark embedding.
Traditional methods design embedding mechanisms in imperceptible spatial\cite{chan2004hiding} or frequency domains\cite{navas2008dwt} to insert watermarks.
Deep learning methods like Hidden \cite{zhu2018hidden} and CIN \cite{ma2022towards} employ encoder-noise-decoder architectures to learn robust watermarking schemes, while techniques like MBRS \cite{jia2021mbrs}, StegaStamp \cite{tancik2020stegastamp}, Pimog \cite{fang2022pimog}, LFM \cite{wengrowski2019light}, and DeNol \cite{fang2023denol} utilize differentiable noise layers during training to simulate real-world distortions—such as JPEG compression, screenshot capture, or photographic degradation—to enhance the robustness of the embedded watermark.
Furthermore, such as Robust-Wide \cite{hu2024robust} designed denoise sampling guidance module and OmniGuard \cite{Zhang_2025_CVPR} proposed a lightweight AIGC editing simulation layer to against AIGC-editing. 

\subsection{In-Generation Image Watermarking}
% \subsubsection{In-Generation watermarking}
% \noindent\textbf{\emph{In-Generation Watermarking.}}
In-generation watermarking refers to embedding watermarks directly during the image creation process, rather than via post-processing.
Previous works such as WatermarkDM \cite{zhao2023recipe}, ProMark \cite{asnani2024promark}, and Diffusion-Shield \cite{cui2023diffusionshield} have embedded copyright watermarks into training datasets to influence dataset attribution, allowing extraction of the watermark from images generated by models trained or fine-tuned on these datasets. Obviously, this technique is inflexible and resource-intensive. 

% Among training-free methods, two primary strategies have emerged: 
Recent research has identified two primary strategies:
\textbf{\emph{Initial Noise Modulation}}. Methods like Tree-Ring \cite{wen2023tree}, GaussMarker \cite{li2025gaussmarker} and Gaussian Shading \cite{yang2024gaussian} embed watermark features by modifying the Fourier spectrum of initial Gaussian noise vectors, or inject encrypted Gaussian-distributed patterns into the initial noise. However, altering the randomness of initial noise distributions can negatively impact both the quality and diversity of generated images, since these changes may reduce the natural appearance and variety of outputs. 
\textbf{\emph{Latent Space Adaptation}}. 
For instance, Stable Signature\cite{fernandez2023stable} fine-tune the VAE decoder for each digital fingerprint, which limits scalability. 
WOUAF\cite{kim2024wouaf} and FSW\cite{xiong2023flexible} embed flexible watermarks by introducing auxiliary networks and fine-tuning the VAE decoder.
RoSteALS \cite{bui2023rosteals} explored the feasibility of leveraging latent space redundancy to embed watermarks without modifying the decoder. 
Similarly, LaWa \cite{rezaei2024lawa} and WMAdapter\cite{ci2024wmadapter} integrate watermark features into latent variables via auxiliary networks while keeping decoder parameters frozen, thus maintaining scalability and efficiency.

The above methods only meet the requirements for watermark extraction. In this paper, GenPTW achieves both watermark extraction and tamper location through single-stage embedding, while maintaining high fidelity and robustness.

% AI生成图像检测 lawa的描述
% Detection of AI-Generated Images. Considering the risks of AIGC, many
%  works focus on the passive detection of generated/manipulated images. These
%  methods are well-studied for deep-fakes using inconsistencies in generated im
% ages [11,23,34] as well as generator traces in the spatial [38,64] or frequency [21,
%  74] domains. However, they have poor performance because they fall behind
%  the rapid evolution of generative models. Similar approaches are proposed for
%  diffusion models [14,50], but they are also shown to suffer from low detection
%  accuracy and high false rate [19].

%% file: sec/3-method.tex
\section{Method}
\begin{figure*}[h]
  \centering
    \includegraphics[width=1.0\textwidth]{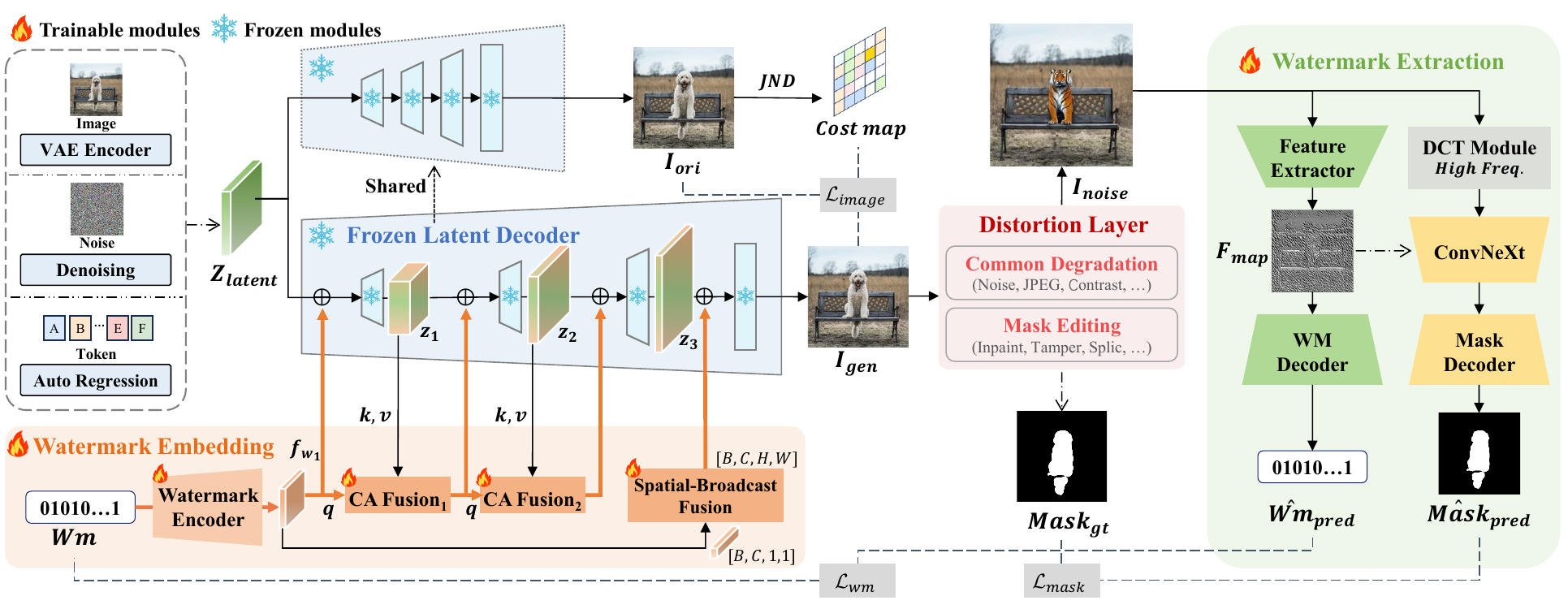}
  % \caption{The Framework of Our Method.}
  \caption{The Framework of \textbf{GenPTW}. 
A Wm plug-in inserted during generation \textbf{without modifying the original model}.}
  \label{main}
\end{figure*}
\subsection{Overall Framework of GenPTW}
% 如图~\ref{main}所示，GenPTW 在潜空间中嵌入水印信息，在统一架构中实现生成归属与篡改定位。
As shown in Fig.~\ref{main}, GenPTW embeds watermark information into the latent space to achieve provenance tracing and tamper localization within a unified design.
% GenPTW 接收任意二进制水印，通过编码后注入生成模型的 latent 解码过程，无需修改原始结构。
Given a binary message, GenPTW encodes it and injects the signal into the latent decoding process without altering the original generative model.
% CAF 根据潜特征语义自适应选择嵌入方式，SF 将水印作为空间提示注入末端特征，以支持精细定位。
CAF adaptively selects embedding strategies based on latent semantics, while SF injects spatial priors into the final decoding layer to support fine-grained localization.
% 在提取阶段，图像经由共享编码器提取特征，用于水印解码与篡改检测两个任务。
In the extraction phase, the image is processed by a shared encoder to extract features for both watermark decoding and tamper localization.
% 共享特征骨干提升任务间的一致性与整体性能。
The shared backbone encourages consistency between tasks and improves joint performance.
% 训练过程中引入扰动模拟层，增强模型对 AIGC 编辑的鲁棒性。
% A distortion simulation layer is added during training to improve robustness against AIGC edits.
To enhance robustness against AIGC edits, we incorporate a distortion simulation layer during training.
% 结合基于代价图的 JND 感知损失，有效约束扰动强度，保障图像质量。
In addition, a JND-aware perceptual loss constrains watermark perturbations through a pixel-wise cost map, effectively preserving visual fidelity.
% 下文将对框架的各个模块进行详细介绍。
Each component of the framework is detailed in the following subsections.

\subsection{Multi-scale Latent Space Embedding}
In general, diffusion models are trained either in the image space or in a compact latent space to reduce computational cost and memory consumption. 
For LDMs, the output of the diffusion process lies in this latent space.
Similarly, VAR models also operate in a latent space to improve modeling efficiency and scalability. 
These models typically incorporate an image autoencoder to map images into a compact latent space, where the image $I_\text{source} \in \mathbb{R}^{H \times W \times 3}$ is encoded into a latent representation $z = \mathcal{E}(I_\text{source}) \in \mathbb{R}^{H/\alpha \times W/\alpha \times C}$ by a factor of $\alpha$, usually $\alpha = 8$ , and decoded by $\mathcal{D}(z)$ in a multi-stage manner. 
% During generation, the diffusion model or VAR synthesizes the latent $Z_{latent}$, which is then progressively upsampled by the decoder to reconstruct the final image.
During generation, the latent representation $\mathbf{Z}_{\text{latent}}$ can be obtained in three different ways: it may be synthesized by a diffusion model, generated by a VAR model, or produced through image compression. The decoder then progressively upsamples $\mathbf{Z}_{\text{latent}}$ to reconstruct the final image.

To embed watermark information, we adopt a coarse-to-fine strategy that injects the message into latent features at multiple decoder stages.  
The decoder performs $\log_2(\alpha)$ upsampling operations to recover the original image resolution, and we embed a scale-specific watermark after each upsampling step.
Given a $k$-bit binary watermark message $Wm \in \{0,1\}^k$, a watermark encoder $E_{wm}$ generates the initial watermark feature $f_{w_1}$ that matches the shape of $z_{latent}$.  
This feature is added to the latent before the first upsampling stage.  
We expect $f_{w_1}$ to evolve alongside the decoding path and progressively inject the watermarking.

\begin{figure}[h]
    \centering
    \includegraphics[width=1.0\linewidth]{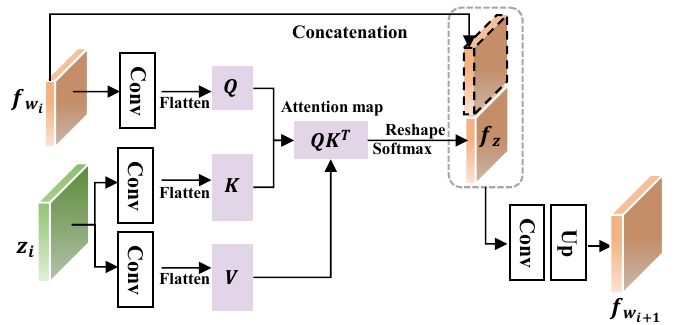}
    \caption{Illustration of Cross-Attention Fusion Block.}
    \label{caf}
\end{figure}

\begin{figure}[h]
    \centering
    \includegraphics[width=0.74\linewidth]{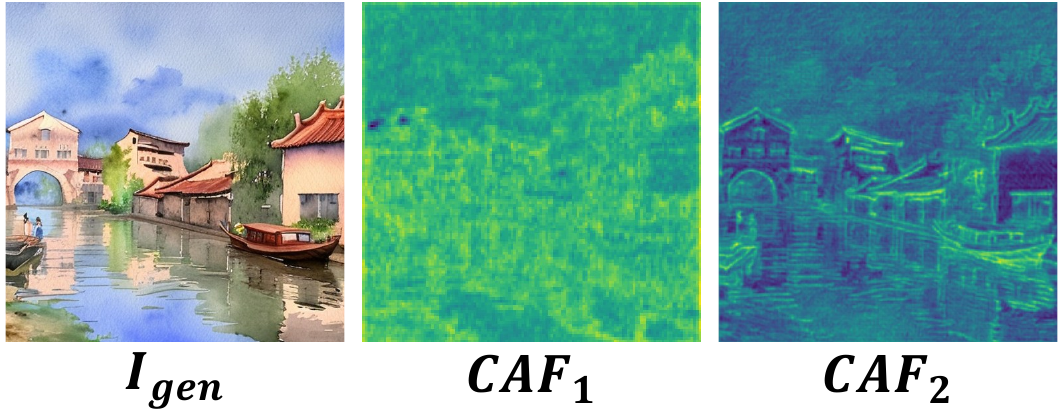}
    \caption{{Average attention maps during image generation with the watermark module.}}
    \label{att}
\end{figure}
\subsubsection{Cross-Attention Fusion Module}
To facilitate the gradual refinement of watermark features during decoding, we employ a Cross-Attention Fusion (CAF) module at each intermediate decoding stage $i \in {1, \ldots, \log_2(\alpha)-1}$.
As illustrated in Fig.~\ref{caf}, the latent representation $z_i$ and the evolving watermark feature $f_{w_i}$ are conceptualized as two interacting modalities. Unlike prior approaches that directly superimpose watermark signals onto feature maps, our design allows the latent representation to actively guide the injection of watermark information. The CAF module generates a residual watermark feature that aligns with the current latent distribution, enabling structurally coherent integration throughout the decoding process—much like ink naturally diffusing along the texture of paper:
\begin{equation}
f_{w_{i+1}} = CAF_{i}(f_{w_i}, z_i)
\end{equation}

\paragraph{CAF Structure.} The CAF block implements a cross-attention mechanism in which $f_{w_i}$ serves as the query, and $z_i$ is used as both the key and value. The inputs are first projected into a shared embedding space via $1 \times 1$ convolutions:

\begin{equation}
% \begin{gathered}
Q = \text{Conv}_Q(f_{w_i}),\
K = \text{Conv}_K(z_i),\
V = \text{Conv}_V(z_i)
% \end{gathered}
\end{equation}

The attention output is computed using scaled dot-product attention:
\begin{equation}
\text{Attention}(Q, K, V) = \text{softmax}\left( \frac{Q K^\top}{\sqrt{d}} \right) V
\end{equation}

The resulting context feature is concatenated with $f_{w_i}$ and passed through a lightweight residual convolutional block to obtain the updated watermark feature:
\begin{equation}
f_{w_{i+1}} = \text{Conv}_{\text{Res}}\left( \left[ \text{Attention}(Q, K, V),\ f_{w_i} \right] \right)
\end{equation}

Finally, the updated watermark feature is added to the current latent representation to form the watermarked latent:
\begin{equation}
z_{wm_i} = z_i + f_{w_{i+1}}
\end{equation}

The fused latent $z_{wm_i}$ is then forwarded to the next decoding stage, allowing the watermark signal to evolve coherently alongside the image features. This interaction mechanism ensures that watermark information is integrated in a latent-aware and transformation-consistent manner, without disrupting the generative pathway.

\subsubsection{Spatial Fusion Module}

At the final stage ($i = \log_2(\alpha)$), the latent representation reaches the full spatial resolution of the image. At this point, most structural and textural information has been reconstructed through the decoder, and the final latent features serve as the immediate precursor to the output image. To inject dense and spatially aligned watermark information at this high-resolution level, we introduce a Spatial Fusion (SF) module. It transforms the initial watermark feature $f_{w_1}$ into a spatial prior and injects it into the final latent representation via feature-wise residual fusion:
\begin{equation}
z_{\text{wm}_{i}} = \text{SF}(f_{w_1}, z_{i})
\end{equation}

\paragraph{SF Structure.}
The SF module first flattens the initial watermark feature $f_{w_1} \in \mathbb{R}^{B \times C' \times H' \times W'}$ into a global vector $\phi \in \mathbb{R}^{B \times D}$, which is then projected to match the channel dimension of the target latent feature through a linear layer:
\begin{equation}
\phi' = \text{Linear}(\text{Flatten}(f_{w_1})) \in \mathbb{R}^{B \times C}
\end{equation}

This projected vector $\phi'$ is broadcast spatially to match the shape of the full-resolution latent feature $z \in \mathbb{R}^{B \times C \times H \times W}$:
\begin{equation}
\Phi = \text{Broadcast}(\phi') \in \mathbb{R}^{B \times C \times H \times W}
\end{equation}

The broadcasted watermark prior $\Phi$ is concatenated with the latent feature $z$ along the channel dimension and passed through a lightweight convolutional fusion block. The result is integrated back into the latent space via residual addition:
\begin{equation}
z_{\text{wm}_{i}} = z + \text{Conv}_{\text{fuse}}(\text{Concat}(z_{i}, \Phi))
\end{equation}

This spatial injection mechanism enables the watermark signal to be embedded explicitly at each pixel location, making the final representation more responsive to localized perturbations. By aligning a globally consistent watermark with the full-resolution latent features, the SF module facilitates fine-grained tamper detection without compromising image reconstruction fidelity.

Finally, the remaining decoding layers convert the watermarked latent representation into the generated image $\mathbf{I}_{\text{gen}}$. This output is then passed through a distortion layer to produce the degraded version $\mathbf{I}_{\text{noise}}$. Detailed configurations are provided in the Appendix.

\subsection{Tamper-Aware Extractor}

To enable robust watermark decoding and accurate tamper localization in a cooperative manner, we propose a Tamper-Aware Extractor that jointly exploits embedded watermark cues and high-frequency features for manipulation analysis. Unlike prior designs that treat watermark extraction and tamper detection as independent tasks, our framework structurally integrates the two objectives into a unified pipeline where they mutually reinforce each other.

Specifically, we employ a shared feature extractor ${E}_{f}$ to process the distorted image $\mathbf{I}_{\text{noise}}$ and generate a shared feature representation $\mathbf{F}_{\text{map}}$. This feature map is then passed to the watermark decoder ${D}_{\text{wm}}$ to recover the embedded watermark message:

\begin{equation}
\begin{gathered}
\mathbf{F}_{\text{map}} = {E}_{f}(\mathbf{I}_{\text{noise}}), \
\hat{\mathbf{Wm_{pred}}} = {D}_{\text{wm}}(\mathbf{F}_{\text{map}})
\end{gathered}
\end{equation}

To improve localization sensitivity to subtle structural inconsistencies, we extract high-frequency priors from $\mathbf{I}_{\text{noise}}$ using the Discrete Cosine Transform (DCT)~\cite{Digital_Image_Processingl_2018}, yielding a high-frequency map $\mathbf{I}_h$ that highlights potential tampering artifacts.

The feature map $\mathbf{F}_{\text{map}}$ is concatenated with $\mathbf{I}_h$ to form the combined input $\mathbf{I}_{\text{all}}$, which is then fed into a ConvNeXt-based multi-scale encoder ${CN}_{\text{Enc}}$ to extract hierarchical semantic features:

\begin{equation}
\mathbf{I}_{\text{all}} = \{\mathbf{I}_h,\ \mathbf{F}_{\text{map}}\}
\end{equation}

\begin{equation}
\begin{gathered}
\{F_{s_1}, F_{s_2}, F_{s_3}, F_{s_4}\}
= {CN}_{\text{Enc}}(\mathbf{I}_{\text{all}}), \\
F_{s_i} \in \mathbb{R}^{\frac{H}{2^{i+1}} \times \frac{W}{2^{i+1}} \times C_i},\ i = 1,2,3,4.
\end{gathered}
\end{equation}

Here, $C_i$ denotes the number of channels at each scale. The multi-level features are subsequently fused and decoded to yield a dense tampering probability map.

To this end, we adopt a hierarchical mask decoder ${D}_{\text{mask}}$ that first upsamples each feature map to a common resolution using transposed convolutions. The aligned features are then concatenated and processed through a lightweight convolutional head to predict the final tampering mask:
\begin{equation}
\hat{\mathbf{Mask}}_{pred} = {D}_{\text{mask}}(\{F_{s_1}, F_{s_2}, F_{s_3}, F_{s_4}\})
\end{equation}

The performance of watermark extraction is measured using binary cross-entropy loss between the predicted watermark $\hat{\mathbf{Wm_\textit{pred}}}$ and the ground-truth message $\mathbf{Wm}$:
% \begin{equation}
%     L_{\text{wm}} = \lambda_{\textit{k}} 
%  \ell_{\text{bce}}(\hat{\mathbf{Wm}}, \mathbf{Wm})
% \end{equation}
\begin{equation}
\mathcal{L}_{\textit{wm}}
= \lambda_{\textit{k}}\,
\ell_{\mathrm{bce}}(\hat{\mathbf{Wm_\textit{pred}}}, \mathbf{Wm})
\end{equation}

For tamper localization, we compute a combination of binary cross-entropy loss and edge-aware loss~\cite{ma2023iml} between the predicted mask $\hat{\mathbf{Mask}}_{\textit{pred}}$ and the ground-truth mask $\mathbf{Mask}_{\textit{gt}}$:
% \begin{align}
% L_{\text{mask}} =\ 
% & \lambda_{\textit{m}} \cdot \ell_{\text{mse}}(\hat{\mathbf{Mask}}_{\text{pred}}, \mathbf{Mask}_{\text{gt}}) \nonumber \\
% & + \gamma \cdot \ell_{\text{edge}}(\hat{\mathbf{Mask}}_{\text{pred}}, \mathbf{Mask}_{\text{gt}})
% \end{align}

\begin{align}
\mathcal{L}_{\textit{mask}}
&= \lambda_{\textit{m}}\, 
\ell_{\mathrm{bce}}(
\hat{\mathbf{Mask}}_{\textit{pred}},\,
\mathbf{Mask}_{\textit{gt}}
) \nonumber \\
&\quad + \gamma\,
\ell_{\mathrm{edge}}(
\hat{\mathbf{Mask}}_{\textit{pred}},\,
\mathbf{Mask}_{\textit{gt}}
)
\end{align}

where $\gamma$ is set to 20.

\subsection{Ensuring Visual Quality}

Our method jointly embeds watermark signals for provenance tracing and tamper localization, which inevitably introduces stronger perturbations than single-task designs. To preserve visual fidelity, we incorporate a Just-Noticeable-Difference (JND)–guided loss to suppress perceptible artifacts.
During training, we generate a clean image $\mathbf{I}_{\text{ori}}$ (without watermark injection) and a watermarked image $\mathbf{I}_{\text{gen}}$ (with injection enabled). The clean image is only used for supervision and perceptual loss computation.

We first compute a map $\mathrm{JND}(\mathbf{I}_{\text{ori}}) \in \mathbb{R}^{3 \times H \times W}$ to estimate pixel-level visibility thresholds, and construct a cost map:
\begin{equation}
\text{Cost Map} = 1 - \alpha_{\mathrm{JND}} \cdot \mathrm{JND}(\mathbf{I}_{\text{ori}})
\end{equation}
The JND-weighted residual loss is defined as:
\begin{equation}
\ell_{\textit{ct}} = \text{Cost Map} \odot \mathbf{I}_{\text{gen}}
\end{equation}
To further constrain distortion, we combine pixel-wise MSE:
\begin{equation}
\ell_{\textit{I}} = \| \mathbf{I}_{\text{gen}} - \mathbf{I}_{\text{ori}} \|_2^2
\end{equation}
with the LPIPS loss~\cite{zhang2018unreasonable}, which better captures perceptual differences. The total visual quality loss is:
\begin{equation}
\mathcal{L}_{\mathrm{image}} = \lambda_{\textit{I}} \ell_{\text{I}} + 
\lambda_{\textit{LPIPS}} \ell_{\textit{LPIPS}} + \lambda_{\textit{ct}} \ell_{\textit{ct}}
\label{eq:image_rec_loss}
\end{equation}

%% file: sec/4-exper.tex
\begin{figure*}[t]
    \centering
    \includegraphics[width=1.0\textwidth]{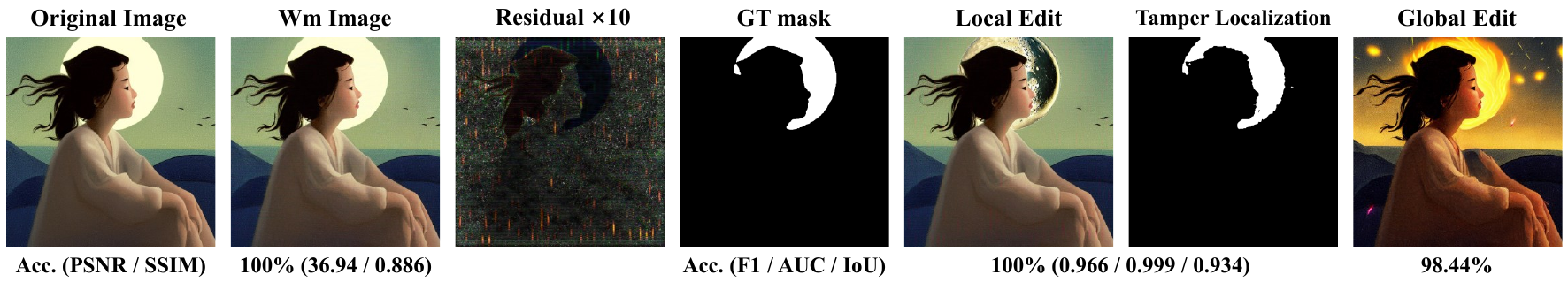}
    \caption{{Qualitative examples of generated images using GenPTW.}}
    \label{lizi}
\end{figure*}
\section{Experiments}

% \begin{figure*}[h]
%   \centering
%     \includegraphics[width=1.0\textwidth]{figure/wm.pdf}
% \caption{\textbf{Qualitative examples of generated images using GenPTW.} }
%   \label{wm}
% \end{figure*}
\begin{table}[t!]
    \centering
    \scriptsize{
    \begin{tabular}{lcccccc}
        \toprule
        & \multicolumn{2}{c}{SD Inp.} & \multicolumn{2}{c}{Splicing} & \multicolumn{2}{c}{Lama}\\
        \cmidrule(lr){2-3} \cmidrule(lr){4-5} \cmidrule(lr){6-7}
        {Method} & F1 & AUC & F1 & AUC & F1 & AUC\\
        \hline
        \multicolumn{7}{c}{\cellcolor[HTML]{EFEFEF}Post-hoc / COCO Test Images} \\ 
        % \hline
        MVSS-Net
        & 0.182 & 0.491 
        & 0.419 & 0.802 
        & 0.027 & 0.508 \\
        CAT-Net 
        & 0.148 & 0.676 
        & 0.200 & 0.722 
        & 0.149 & 0.727 \\
        PSCC-Net
        & 0.170 & 0.504  
        & 0.192 & 0.688 
        & 0.134 & 0.333 \\
        IML-ViT
        & 0.215 & 0.592  
        & 0.469 & 0.759 
        & 0.107 & 0.460 \\
        \hline
        % \hline
        EditGuard
        & {0.964} & 0.973 
        & {0.932} & 0.989 
        & {0.967} & 0.966 \\
        OmniGuard
        & 0.960 & \textbf{0.997}
        & 0.916 & \textbf{0.995}
        & 0.951 & \textbf{0.999} \\
        GenPTW*
        & {0.946} & \textbf{0.997} 
        & {0.930} & {0.994} 
        & 0.956 & {0.998} \\
        \hline
        \multicolumn{7}{c}{\cellcolor[HTML]{EFEFEF}In-generation / AI-generated} \\ 
        % \hline
        % GenPTW^{\dagger}
        % & {0.945} & {0.991} 
        % & {0.906} & {0.899} 
        % & 0.946 & {0.959} \\
        % GenPTW^{\ddagger}
        % & \textbf{0.989} & \textbf{0.999} 
        % & {0.925} & \textbf{0.993} 
        % & 0.955 & {0.996} \\
        GenPTW
        & \textbf{0.973} & {0.996} 
        & \textbf{0.965} & {0.993} 
        & \textbf{0.976} & {0.997} \\
        \bottomrule
    \end{tabular}}
    \caption{{Localization performance of GenPTW and SOTA baselines under clean conditions. GenPTW* is post-hoc; GenPTW is in-generation (SD-based).}}
    \label{localization}
\end{table}

\begin{table*}[t!]
    \centering
    \small
    % \scriptsize
    {
    \begin{tabular}{lll ccc ccc cccc}
        \toprule
        & & &\multicolumn{2}{c}{Global Edit} & \multicolumn{3}{c}{Local Edit} & \multicolumn{4}{c}{Common Degradation} \\
    
        \cmidrule(lr){4-5} \cmidrule(lr){6-8} \cmidrule(lr){9-12}
        {Method} & {PSNR/SSIM} & {LPIPS/SIFID}  & P2P & SDIP+ & SDIP & Lama & RD & Blur & Contr & Bright &JPEG\\
        \hline
        \multicolumn{13}{c}{\cellcolor[HTML]{EFEFEF}Post-hoc Watermarking on COCO Test Images} \\ 
        % \hline
        % \multirow{5}{*}{\rotatebox[origin=c]{90}{\textit{Post}}}
        PIMoG(30)
        & 35.94/0.891 & 0.098/0.045
        & 0.656 & 0.618 
        & 0.913 & 0.949 & 0.954 
        & 0.725 & 0.962 & 0.934 & 0.936 \\
        
        SepMark(30)
        & 31.95/0.879 & 0.112/0.054
        & 0.876 & 0.908
        & 0.951 & 0.952 & 0.948 
        & 0.949 & \textbf{0.987} & 0.955 & 0.996 \\
        
        EditGuard(64)
        & 37.12/0.902 & 0.082/0.012
        & 0.534 & 0.596 
        & 0.969 & 0.971 & 0.960 
        & 0.723 & 0.950 & \textbf{0.984} & 0.938  \\

        OmniGuard(100)
        & 37.21/\textbf{0.912} & \textbf{0.069}/\textbf{0.009}
        & 0.934 & \textbf{0.970} 
        & \textbf{0.997} & \textbf{0.994} & \textbf{0.995}
        & \textbf{0.980} &  \textbf{0.987} & 0.960 &  \textbf{0.998} \\
        
        Robust-Wide(64)
        & \textbf{39.18}/0.905 & 0.097/0.044
        & \textbf{0.976} & 0.956 
        & \textbf{0.997} & 0.968 & 0.981 
        & 0.787 & 0.976 & 0.968 & 0.981 \\

        \hdashline
        
        $\text{WOUAF}^{*}\text{(64)}$
        & 28.51/0.791 & 0.144/0.067
        & 0.543 & 0.564 
        & 0.825 & 0.819 & 0.863
        & 0.951 & 0.928 & 0.943 & 0.962\\
        
        $\text{Lawa}^{*}\text{(48)}$
        & 32.94/0.813 & 0.096/0.023
        & 0.537 & 0.584 
        & 0.851 & 0.833 & 0.879
        & 0.973 & \textbf{1.000} &  \textbf{1.000} &  \textbf{0.998}\\
        
        $\text{GenPTW}^{*}\text{(64)}$
        & \textbf{35.56/0.864} & \textbf{0.077}/\textbf{0.002}
        & \textbf{0.963} & \textbf{0.975} 
        & \textbf{0.998} & \textbf{0.997} & \textbf{0.996} 
        & \textbf{0.999} & \textbf{1.000} & \textbf{1.000} & 0.996\\
        \hline

        \multicolumn{13}{c}{\cellcolor[HTML]{EFEFEF}In-generation Watermarking on AI-Generated} \\  
        % \hline
        
        Stable Signature(48)
        &31.43/0.834 & 0.123/0.064
        & 0.561 & 0.626
        & 0.905 & 0.894 & 0.884
        & 0.784 & 0.914 & 0.943 & 0.914\\
        
        WOUAF(64)
        &30.71/0.847 & 0.130/0.061
        & 0.587 & 0.601 
        & 0.874 & 0.883 & 0.916
        & 0.981 & 0.971 & 0.975 & 0.991\\
        
        Lawa(48)
        &35.14/0.821 & 0.073/0.033
        & 0.591 & 0.629 
        & 0.892 & 0.897 & 0.926
        & {0.999} & \textbf{1.000} & \textbf{1.000} & {0.998}\\
        
        % GenPTW^{\dagger}(64) 
        % &27.56/0.814 &{0.110/0.071}
        % & {0.936} & {0.921} 
        % & {0.961} & {0.945} & {0.939}
        % & {0.998} & {0.976} & \textbf{0.990} & {0.983}\\

        % GenPTW^{\ddagger}(64) 
        % &{35.26/0.873} &{0.071/0.003}
        % &{0.949} & {0.950} 
        % & {0.985} &{0.994}  &\textbf{0.988} 
        % & {1.000} & {0.998} & {0.994} & {0.995}\\
        
        GenPTW(64) 
        &\textbf{39.56/0.892} &\textbf{0.069/0.002}
        & \textbf{0.969} & \textbf{0.974} 
        & \textbf{0.990} & \textbf{0.994} & \textbf{0.977} 
        & \textbf{1.000} & \textbf{1.000} & {0.998} & \textbf{1.000}\\
        
        \bottomrule
    \end{tabular}}
    \caption{\textbf{Fidelity and bit accuracy of GenPTW vs. SOTA baselines.} * denotes post-hoc. + denotes image regeneration via an inpainting model. RD denotes random cropping on $\mathbf{I}_{\text{gen}}$, with cropped regions replaced by $\mathbf{I}_{\text{ori}}$.}
    \label{copyright}
\end{table*}
% Note that “SD Inpaint*” denotes the regeneration from the image via an inpainting model, while “SD Inpaint” ensures that the non-edited regions remain entirely consistent with the original image. 

\subsection{Experimental Setup}
We train our models on the MS COCO dataset~\cite{lin2014microsoft}, using segmentation annotations to generate mask. All images and masks are resized to $512 \times 512$, and editing prompts are fixed as “None”.
The test set includes 5,000 natural images from the COCO validation set and 1,000 AI-generated images synthesized by Stable Diffusion v2 from COCO captions. Editing prompts and masks follow the UltraEdit protocol~\cite{zhao2024ultraeditinstructionbasedfinegrainedimage}.
For VAR-based models, we evaluate 1,000 images, one per class from 0 to 999, each paired with a randomly generated mask.
Since our watermark is embedded in latent space, pairing it with a corresponding autoencoder enables post-hoc usage. We therefore evaluate three pipelines: Stable Diffusion-based generation, VAR-based generation, and a post-hoc variant.
Training is conducted on an NVIDIA A100 GPU using the AdamW optimizer (learning rate $1\times10^{-5}$, batch size 2, gradient accumulation steps 8) with a cosine learning rate schedule. Additional implementation details are provided in the Appendix.
\begin{figure*}[h]
  \centering
    \includegraphics[width=1.0\textwidth]{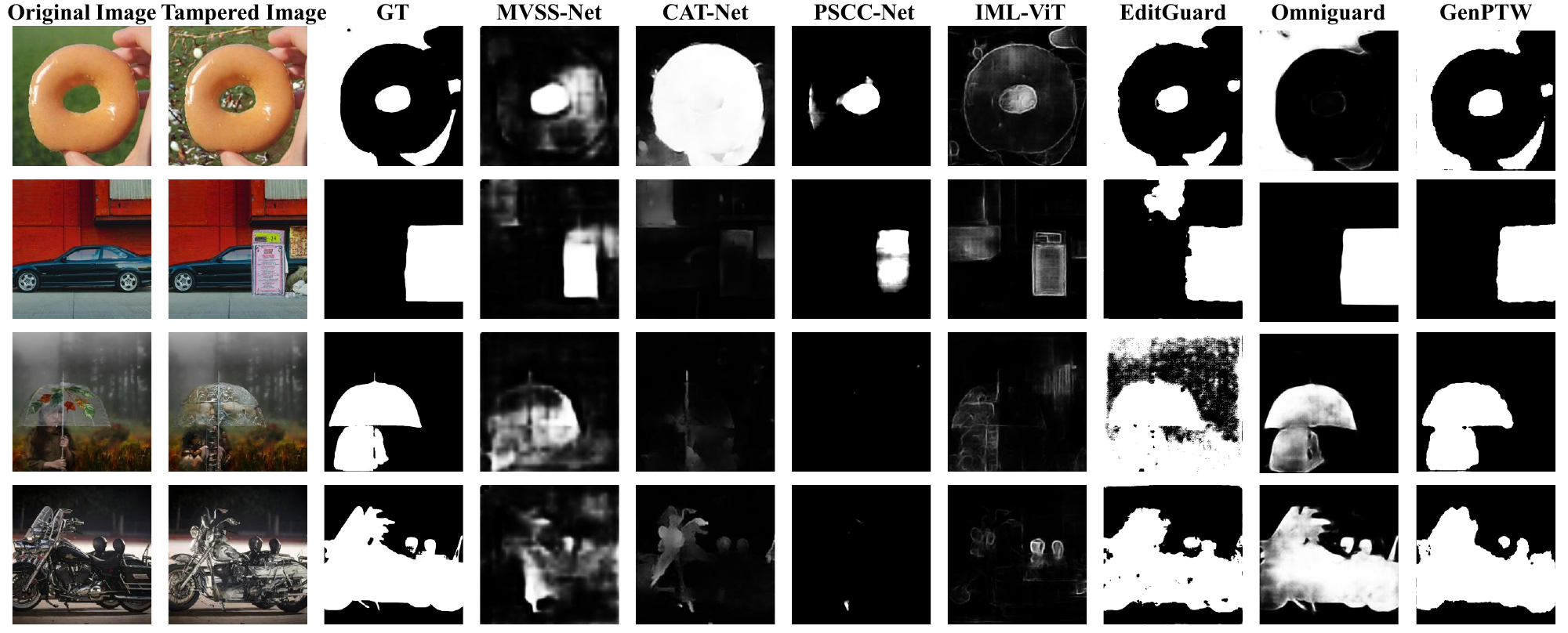}
\caption{{Visualized Comparison between our GenPTW and other methods.}}
  \label{local}
\end{figure*}

\subsection{Comparision with Localization Methods}
To evaluate the tamper localization performance of our proposed \textit{GenPTW}, we compare it against several state-of-the-art passive detection methods, including PSCC-Net~\cite{liu2022pscc}, MVSS-Net~\cite{dong2022mvss}, CAT-Net~\cite{kwon2021cat}, and IML-ViT~\cite{ma2023iml}, as well as proactive watermark-based approaches EditGuard~\cite{zhang2024editguard} and OmniGuard~\cite{Zhang_2025_CVPR}.
We adopt F1 score and AUC as evaluation metrics, with evaluation conducted on the aforementioned test set. 
Manipulations include Stable Diffusion Inpaint~\cite{rombach2022high}, the Lama~\cite{suvorov2022resolution}, and classical splicing, covering both AIGC-based and traditional manipulations.

Figure~\ref{local} compares tamper localization results across methods. Passive approaches like PSCC-Net and IML-ViT often miss manipulations under complex edits, while proactive methods such as EditGuard tend to produce noisy or incomplete masks, with performance sensitive to hyperparameters. In contrast, GenPTW consistently yields accurate and well-aligned masks across diverse perturbations, without requiring post-processing or parameter tuning.
For full-image semantic rewriting tasks like InstructP2P\cite{brooks2023instructpix2pix}, GenPTW remains effective in watermark extraction and tamper detection. Since such edits fundamentally alter global structure, the model often flags the entire image as tampered—reflecting a deliberate design choice to prioritize preserving original visual semantics over adapting to aggressive content shifts.

\subsection{Comparison with Deep Watermarking}
We comprehensively evaluate the performance of GenPTW against both in-generation and post-generation watermarking methods, including Stable Signature, WOUAF, and LaWa for in-generation, and PIMoG~\cite{fang2022pimog}, SepMark~\cite{wu2023sepmark}, EditGuard~\cite{zhang2024editguard}, OmniGuard~\cite{Zhang_2025_CVPR}, and Robust-Wide~\cite{hu2024robust} for post-generation.
All methods are evaluated on 512$\times$512 images from the UltraEdit dataset, except for the 256-resolution VAR-based variant. Test settings follow those described in the experimental setup.

As shown in Table~\ref{copyright}, \textit{GenPTW} consistently achieves higher bit recovery accuracy under various perturbations, with a PSNR of \textbf{39.56}dB—surpassing all in-generation baselines and outperforming several post-generation methods.
Figure~\ref{lizi} presents examples generated by Stable Diffusion v2, followed by both local (SD Inpaint) and global (InstructP2P) edits. Despite significant changes in visual style and structure, GenPTW successfully recovers the embedded watermark, demonstrating strong robustness and generalizability to both global and localized real-world manipulations.
% Furthermore, we evaluate GenPTW on both VAR~\cite{tian2024visual}, DIT~\cite{peebles2023scalable} and SDXL~\cite{podell2023sdxl} architectures. As shown in Table~\ref{vae}, the results demonstrate its generalizability across diverse generative models.
Furthermore, we evaluate GenPTW by training separate models on VAR~\cite{tian2024visual}, DiT~\cite{peebles2023scalable}, and SDXL~\cite{podell2023sdxl} architectures. As shown in Table~\ref{vae}, the results demonstrate its generalizability across diverse generative models.

% 进一步的,我们在VAR和DIT上测试了我们的方案,如表\ref{vae}所示.说明GenPTW的通用性.
        % GenPTW^{\dagger}(64) 
        % &27.56/0.814 &{0.110/0.071}
        % & {0.936} & {0.921} 
        % & {0.961} & {0.945} & {0.939}
        % & {0.998} & {0.976} & \textbf{0.990} & {0.983}\\

        % GenPTW^{\ddagger}(64) 
        % &{35.26/0.873} &{0.071/0.003}
        % &{0.949} & {0.950} 
        % & {0.985} &{0.994}  &\textbf{0.988} 
        % & {1.000} & {0.998} & {0.994} & {0.995}\\
\begin{table}[t!]
    \centering
    \small{
    % \scriptsize{
    % \footnotesize{
    \begin{tabular}{lcccccc}
        \toprule
        % & \multicolumn{2}{c}{SD Inp.} & \multicolumn{2}{c}{Splicing} & \multicolumn{2}{c}{Lama}\\
        {Method} &ACC &PSNR &SSIM & F1 & AUC\\
        \hline
        VAR(256)
        & {0.961} &27.56&0.814 & {0.945} & {0.991} \\
        VAR(512)
        & {0.985} &35.26&0.873 & \textbf{0.989} & \textbf{0.999} \\
        DIT(512)
        & \textbf{0.989} &\textbf{38.48}& \textbf{0.902} & {0.974} & {0.996} \\
        SDXL(1024)
        & {0.981} & {38.01} & {0.897} & {0.957} & {0.995} \\
        \bottomrule
    \end{tabular}}
    \caption{Evaluation on different models under SD Inpaint.}
    \label{vae}
\end{table}

\subsection{Ablation Study}
\subsubsection{Effect of Watermark Embedding Modules}
\begin{table}[t!]
    \centering
    \small{
    \begin{tabular}{lll|ccc}
        \toprule
       $CAF_{{1}}$ & $CAF_{{2}}$ & $SF$
        &ACC. &PSNR &F1 \\
        \hline
         - & - & -
        & {0.594} & {38.87} & {0.520}\\
         \ding{51} & - & -
        & {0.986} & \textbf{40.87} & {0.721}\\
          \ding{51} & \ding{51} & -
        & {0.987} & {39.22} & {0.901}\\
         \ding{51} & \ding{51} & \ding{51}
        & \textbf{0.990} & {39.56} & \textbf{0.973}\\
        \bottomrule
    \end{tabular}}
    \caption{Ablation of Embedding Modules under SD Inpaint.}
    \label{ab1}
\end{table}
We perform ablations by progressively removing different embedding modules. As shown in Table~\ref{ab1}, the SF module has a significant impact on tamper localization, demonstrating its role in enhancing spatial alignment and local sensitivity at high resolution.
Removing the CAF$_2$ module leads to a substantial drop in tamper localization performance, while removing CAF$_1$ primarily affects watermark extraction accuracy. These results indicate that hierarchical embedding is essential for robust watermark propagation.

\subsubsection{Effect of CAF Modules}
\begin{table}[t!]
    \centering
    \small{
    \begin{tabular}{l|lllcccc}
        \toprule
        {Method} &ACC. &PSNR &SSIM &F1 &AUC\\
        \hline
        ADD & \textbf{0.997} & {30.06} & 0.805 &{0.743} &0.596\\
        Fuse & 0.989 & {18.00} & 0.369 &{0.845} &0.739\\
        CAF & \textbf{0.997} & \textbf{35.54} & \textbf{0.863} &\textbf{0.861} &\textbf{0.981}\\
        \bottomrule
    \end{tabular}}
    \caption{Ablation Study of the CAF under SD Inpaint.}
    \label{ab2}
\end{table}
To evaluate the effectiveness of the CAF, we ablate its cross-attention structure and replace the watermark injection with either direct addition (ADD) or convolutional fusion (Fuse). As shown in Table~\ref{ab2}, the ADD variant exhibits poor robustness, while Fuse enhances tamper localization at the expense of visual fidelity. In contrast, the complete CAT module achieves an optimal balance between robustness and perceptual quality. 
Fig.~\ref{att} shows that the two CAFs attend to different structure–texture boundary information; when combined with the evidence in Table~\ref{ab1}, it indicates that $CAF_1$ enhances watermark robustness whereas $CAF_2$ improves localization accuracy.
We provide the overhead analysis in the appendix.
% The attention map illustrated in Fig.~\ref{att} reveals semantically meaningful focus regions, further substantiating the efficacy of the CAT design.
% The attention map illustrated in Fig.~\ref{att} shows that the two CAFs focus on different structure-texture boundary information;with Table 4, CAF₁ enhances watermark robustness and CAF₂ improves localization accuracy.
\subsubsection{Impact of Tamper Localization Inputs}
\begin{table}[t!]
    \centering
    \small{
    \begin{tabular}{lll|ccc}
        \toprule
        $\mathbf{I}_\text{noise}$ & $\mathbf{I}_{h}$ & $\mathbf{F}_{\text{map}}$
        &ACC. &PSNR &F1 \\
        \hline
        \ding{51} & - & -
        & {0.964} & {35.59} & {0.958}\\
        - & \ding{51} & -
        & {0.970} & {36.82} & {0.959}\\
        \ding{51} & - & \ding{51}
        & \textbf{0.999} & {36.77} & {0.962}\\
        - & \ding{51} & \ding{51}
        & {0.991} & \textbf{37.85} & \textbf{0.974}\\
        \bottomrule
    \end{tabular}}
    \caption{Impact of different input configurations for the tamper localization branch under SD Inpaint.}
    \label{ab3}
\end{table}
% We perform an ablation study on the input configurations of the tamper localization branch, as shown in Table~\ref{ab3}. Using either high-frequency image $I_h$ or $I_{noise}$ alone yields comparable performance, while incorporating the fused feature $F_{map}$ significantly improves both localization and decoding accuracy. The combination of $I_h$ and $F_{map}$ achieves the best results, validating the effectiveness of our joint optimization design.
We conduct an ablation study on the input configurations of the tamper localization branch, as shown in Table~\ref{ab3}. Using either the high-frequency image $\mathbf{I}_h$ or $\mathbf{I}_\text{noise}$ alone achieves acceptable performance, while incorporating the fused feature $\mathbf{F}_\text{map}$ significantly improves both tamper localization and watermark extraction.
% Combining $I_h$ with $F_{map}$ yields the best results, validating the effectiveness of our joint optimization design.
The $\mathbf{I}_h + \mathbf{F}_\text{map}$ combination achieves accuracy comparable to $\mathbf{I}_\text{noise} + \mathbf{F}_\text{map}$ while providing the best visual quality, yielding the optimal trade-off.